# Berry Curvature Enhanced Nonlinear Photo Response of Type-II Weyl Semimetals

Junchao Ma,[1] Qiangqiang Gu,[1] Yinan Liu,[1] Jiawei Lai,[1] Yu Peng,[2] Xiao Zhuo,[1] Zheng Liu,[2] Jian-Hao Chen,[1,3]† Ji Feng,[1,3]† Dong Sun[1,3]†

[1]International Center for Quantum Materials, School of Physics, Peking University, Beijing 100871, P. R. China
[2]School of Materials Science and Engineering, Nanyang Technological University, 639798, Singapore
[3]Collaborative Innovation Center of Quantum Matter, Beijing 100871, China.

†Corresponding author. Email: sundong@pku.edu.cn (D.S.); jfeng11@pku.edu.cn (J.F.); chenjianhao@pku.edu.cn (J.-H.C.);

**Abstract:**
The experimental manifestation of topological effects in bulk materials under ambient conditions, especially those with practical applications, has attracted enormous research interest. Recent discovery of Weyl semimetal provides an ideal material platform for such endeavors. The Berry curvature in a Weyl semimetal becomes singular at the Weyl node, creating an effective magnetic monopole in the *k*-space. A pair of Weyl nodes carry quantized effective magnetic charges with opposite signs, and therefore, opposite chirality. Although Weyl-point-related signatures such as chiral anomaly and non-closing surface Fermi arcs have been detected through transport and ARPES measurements, direct experimental evidence of the effective *k*-space monopole of the Weyl nodes has so far been lacking. In this work, signatures of the singular topology in a type-II Weyl semimetal $TaIrTe_4$ is revealed in the photo responses, which are shown to be directly related to the divergence of Berry curvature. As a result of the divergence of Berry curvature at the Weyl nodes, $TaIrTe_4$ exhibits unusually large photo responsivity of 130.2 mA/W with 4-μm excitation in an unbiased field effect transistor at room temperature arising from the third-order nonlinear optical response. The room temperature mid-IR responsivity is approaching the performance of commercial HgCdTe detector operating at low temperature, making Type-II Weyl semimetal $TaIrTe_4$ of practical importance in terms of photo sensing and solar energy harvesting. Furthermore, the circularly polarized galvanic response is also enhanced at 4-μm, possibly due to the same Berry curvature singularity enhancement as the shift current. Considering the optical selection rule of Weyl cones with opposite chirality, it may open new experimental possibilities for studying and controlling the chiral polarization of Weyl Fermions through an in-plane DC electric field in addition to the optical helicities.

**Introduction:**

The exploration of the topological effects with useful and unique applications in condensed matter materials attracts a lot of current interest[1-4]. The quantum Hall system is a patent manifestation of topological effect in insulating phase, which, apart from its fundamental significance, has found practical application as metrological standard for resistance. Recently discovered Weyl semimetals (WSM), belonging to the class of topological metallic phases, provide an ideal platform to explore exotic physical effects that relates to topology in gapless materials[5-11]. The surface Fermi arcs and chiral anomaly, characteristics of Weyl Fermions (WFs), have been widely used for experimental verification of WSMs through angle-resolved photoemission spectroscopy (ARPES) [6,7,12,13] and transport measurement[14-16], respectively. The defining feature of a WSM is the divergence of Berry curvature at the Weyl nodes (WNs), leading to topological semi-metallic phases. The divergence behaves like magnetic monopoles of the momentum space and the sign of the monopole determines the chirality. On the other hand, nonlinear optoelectronic responses play a crucial role not only in optical devices but also in probing fundamental properties of quantum materials. Although a nonlinear optical process that involves direct optical transitions to high energy excited states may not capture the low energy singularity at the WNs directly, its measurable effect, such as photo current response, will involve the ground and low energy excited states that are characterized by the geometric and topological nature of the Bloch wave functions at the WNs. These ground and low energy excited states properties of WNs involve the singularity of Berry curvature, an intrinsic geometric property of the Bloch wavefunctions, which plays an essential role in transport phenomena of photoexcited carriers. It has been proposed theoretically that the singular Berry curvatures at the WN could be manifested in macroscopic nonlinear optical response such as second harmonic generation and photo current response through the shift vectors [17-19]. The shift current is the photocurrent generated due to a change in the center of mass upon optical excitation [20,21]. It features ultrafast response and is considered as a promising candidate responsible for the high-efficiency photovoltaic current in solar cells without the p-n junction [22-26]. The shift current response is related to the geometric effects associated with the Berry connection and curvature of the Bloch bands, which in turn form the basis for defining topology in band structures[17,27]. Furthermore, theoretical investigation shows stark contrast in the low frequency dependence of the shift current response of Type-I and Type-II WSM: the shift current diverges in the low frequency $\omega \to 0$ limit for a Type-II WSM, but vanishes ($\propto \omega$) for a Type-I WSM with zero doping [27]. In this work, we experimentally demonstrate the manifestation of the divergent Berry curvature of WNs in the giant mid-infrared photo response from TaIrTe$_4$, a Type-II WSM, largely a result of the shift current generation in a third-order nonlinear optical process. Taking advantage of the giant shift current response, we further demonstrate that the third order nonlinear injection current terms may provide additional control for optical injection of chiral carriers in the Weyl cones through an in-plane electric field.

Our experiment is based on the layered orthorhombic ternary compound tantalum iridium tetratelluride (TaIrTe$_4$), which hosts four well-separated Type-II WNs, the minimal number of WNs imposed by symmetry in a WSM without inversion symmetry. It offers a "hydrogen atom" example of an inversion-breaking WSMs[28]. In addition, TaIrTe$_4$ also hosts larger Fermi arc surface states, and inherits all advantages as a layered material, such as the ability to form van der Waals heterostructures with other 2D layered materials [29]. The $k_c = 0$ cross-section of the Brillion zone harboring all four Weyl points (WPs) is shown in Fig. 1a, along with 3-dimensional band structure around the four WNs. The four Weyl cones can be mapped exactly onto each other through mirror and time reversal symmetry. In the $k_a$ direction, the bands are strongly tilted to form Type-II Weyl cones, while along the $k_b$ direction, the bands are almost upright (Supplementary Fig. S1). The theoretically predicted Fermi level lies about 80 meV below the WNs. For this reason, the WNs are not conveniently accessible in conventional transport and ARPES measurement without special experimental arrangement to tune the Fermi level over extremely large energy range[13]. However, with suitable photon energy through photogalvanic effect measurement, the Weyl properties are demonstrated to be accessible by the photon probe in this work [30].

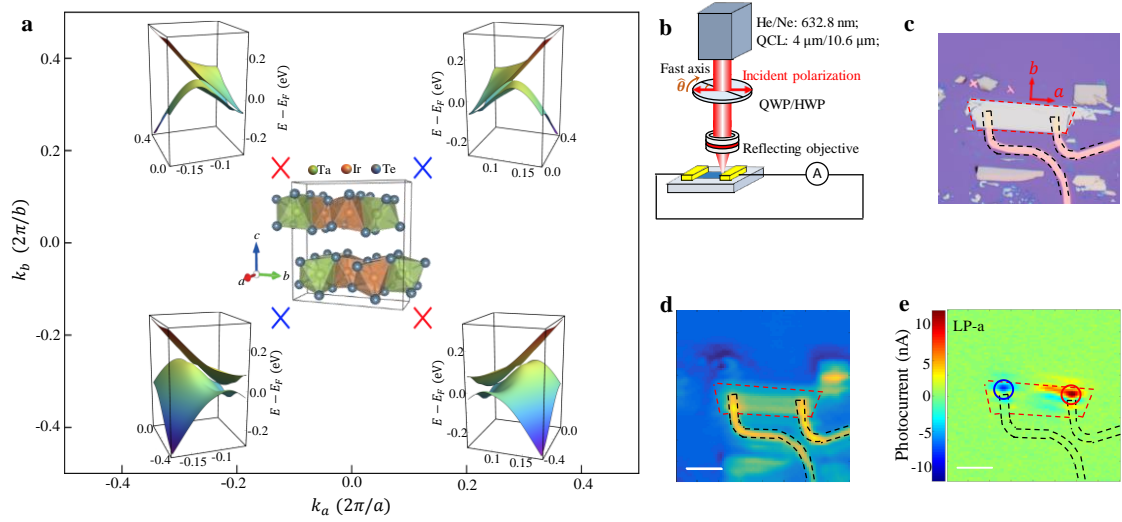

**Fig. 1 | Type-II Weyl Semimetal TaIrTe$_4$ and Scanning Photocurrent Response. a**, TaIrTe$_4$ crystal and band structures. The central inset shows the crystal structure of TaIrTe$_4$. A cross-section of Brillouin zone with $k_c = 0$ is shown. Four crosses indicate the positions of WNs in Brillouin zone, with the red and blue colors corresponding to opposite chirality of the WNs. The four insets display the band structures in the vicinity of the WNs. **b**, Schematic diagram of the polarization resolved scanning photocurrent measurement setup. **c**, Optical microscope image of the device. The red arrows mark the crystallographic *a*- and *b*-axes. **d**, **e**, Scanning reflection and photocurrent images of the TaIrTe$_4$ device at room temperature with 4-μm linear polarized light along crystallographic *a*-axis. The excitation power is 490 nW. The maximum photocurrent response is 12.45 nA and -8.49 nA. All scale bars are 20 μm.

The schematic diagram of the photo current measurement is illustrated in Fig. 1b: continuous wave light sources with different wavelengths are polarization controlled and focused on specific position of a lateral metal-TaIrTe$_4$-metal device (Fig. 1c). Throughout the paper, we have normal incidence of light and we assign the direction of propagation of light as $-\hat{c}$. The scanning reflection and photocurrent (PC) images excited with 4-μm linear polarization along the crystallographic *a*-axis (LP-a) are shown in Fig. 1d and 1e, respectively. The response pattern is qualitatively the same with higher spatial resolution scanning image at 633 nm (Supplementary Fig. S2) and the results indicate the photo response is not only limited to the interface between metal contacts but also extend to regions that are far away from the contact. The scanning photocurrent response pattern is determined by multiple factors such as thickness, doping profile and field line distribution between the contacts, which is a result of the interplay of multiple photocurrent generation mechanisms.

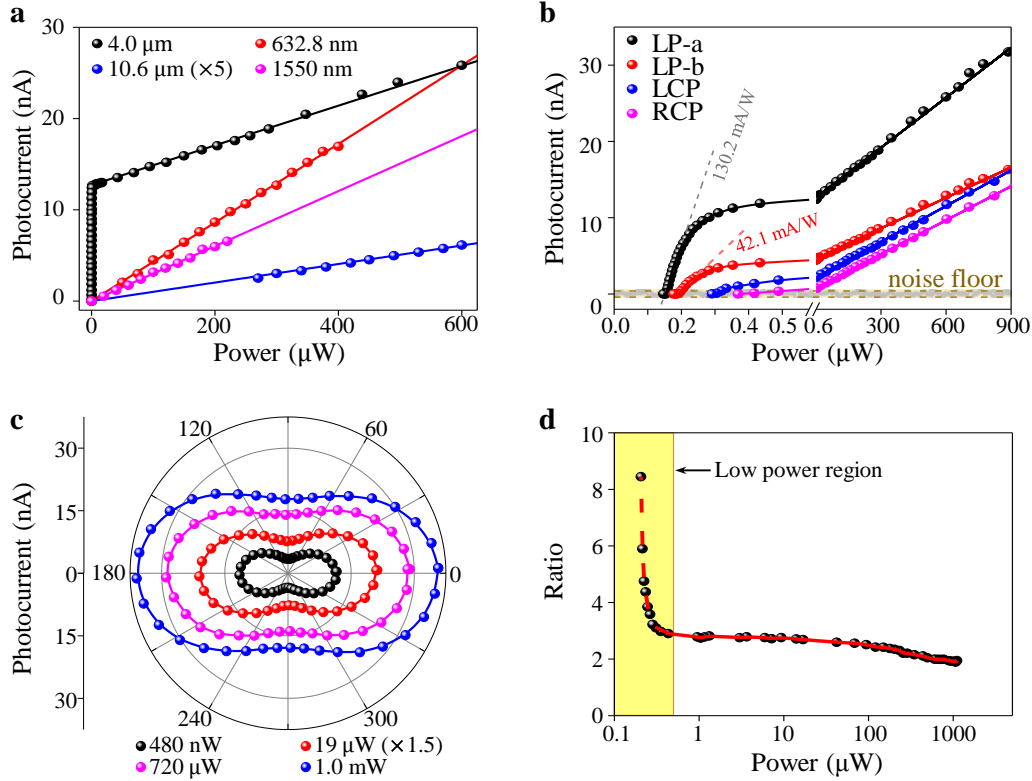

**Fig. 2 | Giant and Anisotropic Shift Current Response**. **a,** Excitation power-dependence of PC response at different wavelengths at fixed excitation spot marked by red circle in Fig. 1e. All the excitations are linearly polarized. The responsivities for 632.8 nm, 1550 nm and 10.6 μm excitations are 43.05 μA/W, 30.13 μA/W, and 2.03 μA/W, respectively. **b,** Excitation power-dependence of the PC response under 4-μm LP-a, LP-b, RCP and LCP excitations, respectively. The gray and red dashed lines show the linear fitting of responsivity of 130.2 mA/W and 42.1 mA/W of linear regions at low excitation power with LP-a (from 155.33 nW to 206.1 nW) and LP-b excitation (192.35 nW to 246.54 nW) respectively. The solid lines are guide to the eye. The noise floor is marked with the shadow region from -0.4 nA to 0.4 nA. **c,** Anisotropic PC response under different excitation power at 4-μm shown in the polar plots. The incident linear polarized light before passing through the wave plate is along the crystallographic *a*-axis. The polar angle is the polarization angle of the linear polarized light with respect to the crystallographic *a*-axis after passing through the half-wave plate. **d,** The excitation power dependence of the anisotropy ratio of the PC ellipse down to the excitation power that has non-zero response along the crystallographic *b*-axis. The *x*-axis is plotted in log scale with the low power region highlighted. Ratio at low power region is marked with a dash line.

Figure 2a shows the power dependence of photoresponse for different excitation wavelengths. While the photo responses at 633 nm, 1.5 μm and 10.6 μm are linear with excitation power over the whole power range, the response at 4 μm is strikingly different in the low power region. In comparison with the responsivity of 22.4 μA/W at high power region, the responsivity at low power region is more than three orders of magnitude larger, reaching a maximum slope of 130.2 mA/W as shown in Fig. 2b. Here we note that there is a "turn on" threshold at low excitation power (~ 155 nW for LP-a and ~192 nW for LP-b), which corresponds to the minimum excitation power to circumvent the potential barriers in the measurement system so that a measureable photocurrent is induced. Such "turn on" threshold is related to interplay of multiple potential barriers in the measurement circuit and cannot be removed by simply adding a fixed bias during the measurement before reaching the damage threshold of the device. Figure 2b further shows the power dependence for different excitation polarizations at 4 μm. The response has obvious polarization dependence. Firstly, with LP-a excitation, the photo-response is much larger than that with linear polarization along crystallographic *b*-axis (LP-b). The response is highly anisotropic, and the long axis of the ellipse of anisotropic response is consistent with the crystallographic a-axis (Fig. 2c). The anisotropy ratio, which is defined as the ratio of photocurrent response with LP-a excitation to LP-b excitation, increases monotonically as excitation power decreases (Fig. 2d). Due to the limit imposed by the "turn on" threshold, the plot of Fig. 2d is cut at the power when the response of LP-b excitation reaches the noise floor (~ 200 nW), also the anisotropy ratio at low powers is biased by the "turn on" threshold of the device. On the other hand, the response to the left circular polarization (LCP) excitation is larger than that to the right circular polarization (RCP) excitation, which corresponds to circular photogalvanic effect (CPGE).

The giant photo responsivity at 4 μm is attributable to the shift current response, which in turn can arise from the diverging Berry curvature at the WNs as predicted by recent theory [17,18,27]. Here, an analysis of the nonlinear photocurrent tensor σ$_{shift}$ that is responsible for the shift current generation is offered in accordance with the crystal symmetry. An approximate calculation of the shift current tensor in TaIrTe$_4$ affords a semi-quantitative analysis of the experimentally observed giant photo-response. As will be shown, the computed σ$_{shift}$ does involve significant contribution from the divergence of geometric quantities near the WNs.

It has been demonstrated experimentally that the shift current response from a Type-I Weyl cone results in a Glass coefficient which is an order of magnitude larger than any other previously measured values [18]. Interestingly, the observed giant shift current response in TaIrTe$_4$ is a result of third order nonlinear effect. The rank-3 tensor $\sigma_{(2)}^{\alpha\beta\gamma}$ for in-plane DC optical response vanishes for $\alpha,\beta = a,b$ (in the plane of the device and perpendicular to the direction of incident laser) in the C$_{2v}$ crystallographic point group. The rank-3 tensor $\sigma_{(2)}^{\alpha\beta\gamma}$ for DC optical response along the *c*-axis is not zero by symmetry; however, experimentally, such optical response is not collected efficiently by the electrodes along the *a*-axis in our experimental geometry as discussed in Supplementary Information IV. The rank-4 tensors $\sigma_{(3)}^{\alpha\lambda\beta\gamma}$ that are responsible for third order nonlinear DC optical response have finite elements involving only planar components. The third order nonlinear DC response necessarily involves a DC electric field *E*$_{DC}$. Experimentally, *E*$_{DC}$ can be provided by multiple effects, such as built-in electric field due to work function difference between TaIrTe$_4$ and the metal contacts or photo-thermoelectric (PTE) field after photoexcitation[31,32]. The observed DC photocurrent is therefore seen as a result of the combination of an in-plane DC field $E_{DC}$ arising from space charge (presumably due to the device geometry) and the optical excitation, in a third order nonlinear process, i.e., $\sigma_{(3)}^{\alpha\lambda\beta\gamma}(0,\omega,-\omega)E_{DC}^{\lambda}E_{opt}^{\beta}(\omega)E_{opt}^{\gamma}(-\omega) + c.c.$

In our approximate calculations, the $E_{DC}$ is treated as an accelerating the Bloch electrons whence displacing the Fermi distribution (see Eq. (2) in Methods). This accelerating DC field then is seen to place the system in a non-equilibrium state without full $C_{2v}$ symmetry, leading to non-zero second-order DC optical response. Thus, the rank-4 tensor is approximated by the second-order DC response with a displaced distribution due to $E_{DC}$.

As seen in Eqs. (3-5), the shift current is an integral of the gauge-invariant quantity $r_{mn}^{\beta}r_{nm;\alpha}^{\gamma} + r_{mn}^{\gamma}r_{nm;\alpha}^{\beta}$ weighted by occupation factors and usual energy denominator (Supplementary Information VI). Thus the non-linear shift current depends on the non-Abelian Berry connection $r_{mn}(k) = \langle u_m(k)|i\nabla_k|u_n(k)\rangle$ with $u_m(k)$ and $u_n(k)$ being the m- and n-th eigenstates of the Bloch Hamiltonian at crystal momentum $k$. As the WNs are monopoles of Berry curvature, the contribution to shift current is expected to be singularly large near WNs. This analysis is based upon the same physical processes invoked in the recent report of the largest second harmonic generation with visible light in TaAs[33].

Our density-functional theory (DFT) calculations (see Supplementary Information VI) enable us to calculate the Berry-phase related quantities such as the Berry connections, from which the third order optical response [21] is computed, within the approximation outlined above. In accordance with the device geometry and sample orientation, $E_{DC}$ is taken to be along the crystallographic *a*-direction in the simulation, however we note the $E_{DC}$ can be along any other in-plane directions, for example the crystallographic *b*-direction as discussed in Supplementary Information VI, and the analysis thereafter stay qualitatively the same. There are four non-zero planar components of rank-4 tensors for $E_{DC}$ along the crystallographic *a*-direction, $\sigma_{(3)}^{aaaa}$, $\sigma_{(3)}^{aabb}$, $\sigma_{(3)}^{baab}$ and $\sigma_{(3)}^{baba}$. The four superscripts, in the order written, correspond respectively to the direction of current, direction of $E_{DC}$, the two optical field directions.

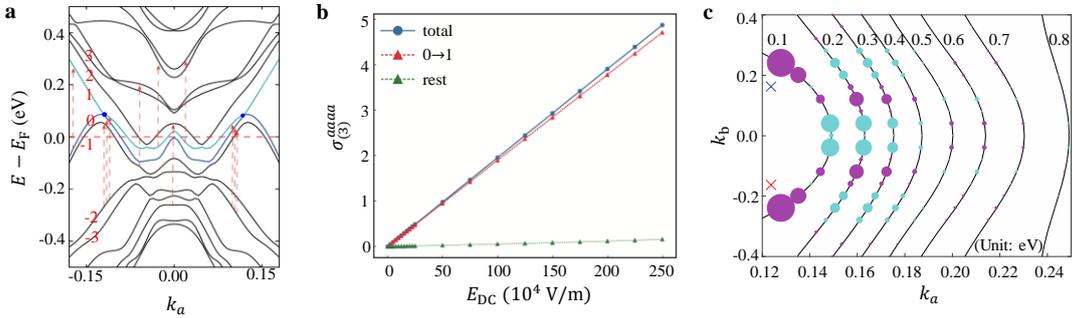

**Fig. 3 | Numerical Simulation of the Shift Current Response. a,** Band structure cutting through a pair of WNs along $k_a$ direction. The red dashed lines indicate theoretically predicted Fermi level and the red arrows indicate the possible optical transitions with 4-μm excitation. The numbers represent the band indices and the 0th-band and 1st-band cross each other to form tilted Weyl cones. **b,** $\sigma_{(3)}^{aaaa}$ as a function of $E_{DC}$ with 4-μm optical excitation at 300K. The solid blue line represents

total $\sigma_{(3)}^{aaaa}$ that consists of all the possible optical transitions. The dashed red line marks the contribution from optical transition across the WN (0->1 band) and the dashed green line is the total contribution from all other transitions. **c,** Fix energy contour plot of matrix elements $r_{mn}^a r_{nm;a}^a + r_{mn}^a r_{nm;a}^a$, which is the integrand of effective third-order optical conductivity tensor at fixed energy contour. The radius of the disc indicates the numerical value of matrix elements and the purple (cyan) color marks the positive (negative) sign. The red (+1) and blue (-1) crosses are the WPs with opposite chirality.

Under our experimental configuration, $\sigma_{(3)}^{aaaa}$ mainly accounts for the shift current response that is measured with LP-a excitation. The 4-μm photon excitation actually involves transition between many bands besides the two forming the Weyl cone. The representative transitions are marked in the band diagram shown in Fig. 3a. Figure 3b further shows that the contribution of $\sigma_{(3)}^{aaaa}$ from individual optical transition across the WN takes the major portion of the $\sigma_{(3)}^{aaaa}$ response under different $E_{DC}$. The WNs are characterized by divergent Berry curvature, which can be thought of the magnetic monopole in the *k*-space. Remarkably, Berry curvature is also the curl of Berry connection, which, along with its derivatives, determine the magnitude of the non-linear optical processes responsible for the shift current. It is therefore expected that, symmetry permitting, the Weyl Fermions can generate significant nonlinear optical response in the form of shift current, as the shift current can enjoy the same boost as the Berry curvature in the vicinity of the Weyl nodes (Fig. 3c). Indeed, after separating the photocurrent response of all optical transitions in the calculations, we found the photocurrent response from 4-μm transition at the Weyl cone (from band 0->1) accounts for over 97 % photocurrent response, while the rest of transitions account for less than 3 %.

According to the above shift current response scheme, the saturation behavior observed at relatively high excitation power (200~500 nW) with 4-μm LP-a excitation (Fig. 2a) is a result of band filling effect in Weyl cones as discussed theoretically in one-dimensional N-site fermion chain model[17]. Specifically, the saturation is determined by the competition of photo absorption rate ($W_{VC}$) and the relaxation mechanism (characterized by a relaxation rate $\Gamma$). If $W_{VC} << \Gamma$, which applies at low excitation region, the shift current is proportional to the intensity as expected, but at high excitation region, $W_{VC} >>\sim \Gamma$, the shift current quickly saturates and becomes intensity independent. The photocurrent rises very quickly at low excitation power (Fig. 2b), where the shift current response dominates. The onset of linear power dependence of the photo response with low responsivity is around 600 nW, where other photo response mechanisms such as PTE dominate. We further note the anisotropy ratio of 1.92 at high excitation power is in good agreement with the recently reported anisotropy ratio determined using DC conductivity measurement[34].

Besides unusual behavior on the excitation power, the wavelength dependence of photo response provides additional support to the Weyl-nature of the shift current responses. As the divergence of Berry curvature occurs in the vicinity of the Weyl cones, the responsivities at large excitation photon energies (1.55 μm and 633 nm) are normal as the transitions are far away from the WNs (Fig. 2a). This supports that the unusual low-power responsivity observed at 4 μm is related to the singularity of the WNs. However, for 10.6-μm excitation, the optical transitions are more close to the WNs compared to that of 4 μm, but the responsivity at low excitation power is normal instead of showing unusual behavior similar to 4 μm. This is because the transitions for 10.6-μm in the vicinity of Weyl cones are completely Pauli blocked and are allowed only due to the temperature smearing the Fermi surface as temperature elevates. As illustrated in Supplementary Fig. S4b, the joint density of state (JDOS) of the 10.6-μm transitions is much lower compared to the 4-μm transitions. On the other hand, with finite temperature $T$ and doping $\mu$, the divergence of $\sigma_{shift}$ as a function of $\omega$ shifts from the WN and is broadened from singularity by energy scale of $k_BT/\hbar$ and $2\mu/\hbar$ [27]. As shown in Supplementary Fig. S4c, the tensor $\sigma_{(3)}^{aaaa}$ accounts for the shift current response at 4-μm transitions and reaches maximum when the Fermi level is 90 meV below the WN, which matches the theoretically calculated Fermi level and is within the range of the experimentally measured doping level of TaIrTe$_4$ sample from the same source through angle resolved photoemission measurement[13]. On the other hand, due to the Pauli blocking, the shift current response is much smaller for 10.6-μm transitions compared to 4-μm transitions in the vicinity of Weyl cones (Supplementary Fig. S4c).

Furthermore, the generated giant shift current from the Weyl cones reflects enhanced in-plane $E_{DC}$, which could further play a role in the injection current response, through other third order nonlinear tensors $\eta_{(3)}^{baab}$ and $\eta_{(3)}^{baba}$ (see Supplementary Information IV). Experimentally, we can separate out the injection current response from the shift current response through the CPGE measurement by rotating a quarter-wave plate as described in the methods session. The excitation optical field $\boldsymbol{E}_{opt}(\hat{\boldsymbol{\theta}})$ corresponds to an LP-a light field passing through a rotating quarter-wave plate with fast axis $\hat{\theta}$ respect to crystallographic $a$-axis ($\hat{a}$), and incidents to the device along crystallographic $c$-axis ($\hat{c}$):

$$\boldsymbol{E}_{opt}(\hat{\boldsymbol{\theta}}) = E_{opt}(\hat{\boldsymbol{a}})\{(\hat{\boldsymbol{\theta}} \cdot \hat{\boldsymbol{a}})\hat{\boldsymbol{\theta}} + i[(\hat{\boldsymbol{c}} \times \hat{\boldsymbol{\theta}}) \cdot \hat{\boldsymbol{a}}]\hat{\boldsymbol{c}} \times \hat{\boldsymbol{\theta}}\} = E_{opt}(\hat{\boldsymbol{a}})(cos\theta\hat{\boldsymbol{\theta}} - isin\theta\hat{\boldsymbol{c}} \times \hat{\boldsymbol{\theta}}) \quad (1)$$

The optical field is circularly polarized for $\theta=\pi/4$ and $3\pi/4$. As described in Supplementary Information VII, the response from the shift tensors is dominated by the $\hat{\theta}$-independent and the $\pi/2$-periodicity parts, while the response from the injection tensors contains only $\pi$-periodicity (CPGE) part, which results in different responsivity under RCP and LCP excitations (Fig. 2b).

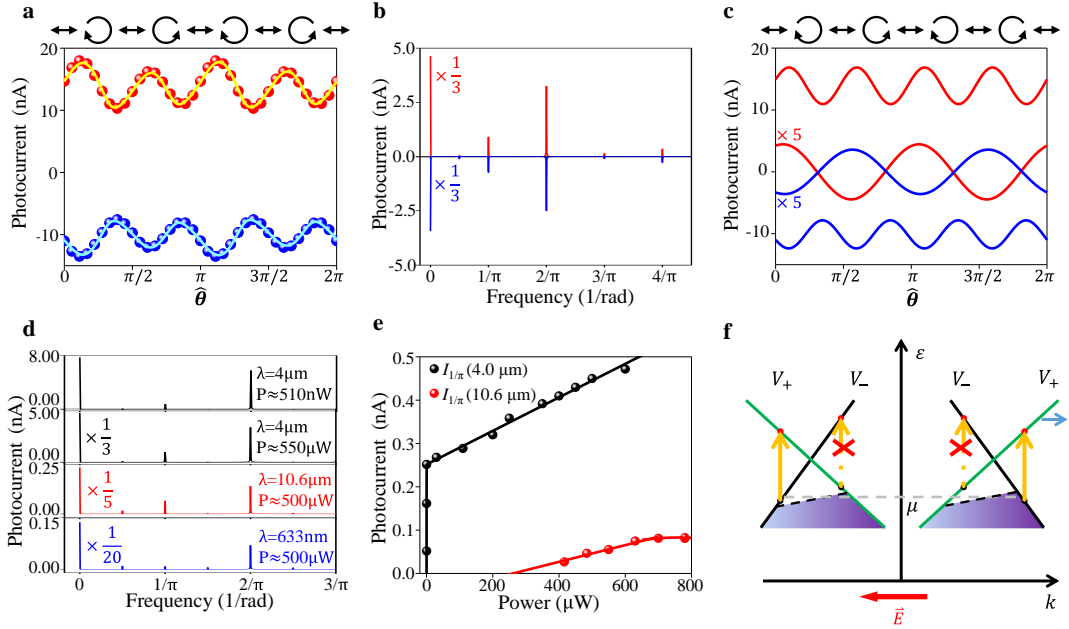

**Fig. 4 | Circular Photogalvanic Response. a,** PC response as a function of $\hat{\theta}$, the angle of the fast axis of quarter-wave plate with respect to the polarization orientation of the incident light. The linear polarization direction of the incident light before passing through the quarter-wave plate has a 60-degree angle with respect to the crystallographic *a*-axis which accounts for the shift of the first maximum (red) or minimum (blue) from $\hat{\theta} = 0$. The measurement is performed on the positive (red dots) and negative (blue dots) photoresponse parts as marked by the red and blue circles in Fig. 1e. The excitation power is 570 μW at 4 μm. **b,** Fourier transform from $\hat{\theta}$ space to the angular frequency space. **c,** π-periodicity ($I_{1/\pi}$) and π/2-periodicity ($I_{2/\pi}$) components from CPGE measurement of positive (red lines) and negative (blue lines) photoresponse positions. **d,** Fourier transform spectrum of CPGE measurement with different wavelengths and excitation powers. **e,** Power dependence of $I_{1/\pi}$ under 10.6 μm and 4.0 μm excitations, respectively. **f,** Schematics of chiral selection rule and CPGE response from a pair of Weyl cones.

The CPGE measurement results are shown in Fig. 4. Figure 4a shows typical results when the light spot is fixed on a negative (blue) and a positive (red) photo response part, respectively. The directions of the net photocurrent are opposite at these two spots, as a result, the in-plane electric field $E_{DC}$ should have opposite directions. The photo response shows obvious but complicated polarization dependence when tuning $\hat{\theta}$ continuously. The Fourier transform from $\hat{\theta}$ space to the angular frequency space (Fig. 4b) can separate components with different angular periodicities. Apart from the low-frequency component that corresponds to the polarization independent photocurrent response ($I_\infty$), two sharp peaks at angular frequencies of 1/π ($I_{1/\pi}$) and 2/π ($I_{2/\pi}$) are observed. The intensity of the minor peak at angular frequency of 1/2π ($I_{1/2\pi}$) represents the noise floor of the photocurrent measurement when rotating the quarter-wave plate (λ/4) over a 2π period. In Fig. 4c, we further show the frequency-filtered photocurrent of different periodicities. If we add the $I_\infty$, $I_{2/\pi}$ and $I_{1/\pi}$ together, it recovers the experimentally measured signal shown by the solid lines in Fig. 4a. Here, $I_{2/\pi}$ corresponds to the in-plane anisotropic response and $I_{1/\pi}$ is the circular polarization dependent component that is responsible for the injection current. Figure 4d further shows the wavelength dependent CPGE measurements, where $I_\infty$ and $I_{2/\pi}$ components are both notable at all excitation wavelengths and the $I_{1/\pi}$ component is negligible at high-energy 633-nm photon excitation. Both 4-μm and 10.6-μm excitations result in significant $I_{1/\pi}$, however, according to the power dependence measurements of $I_{1/\pi}$ shown in Fig. 4e, $I_{1/\pi}$ has cut off at low power with 10.6-μm excitation, while it survives at very low excitation intensity down to the detection limit with 4–μm excitation. The survival of $I_{1/\pi}$ at low excitation power at 4–μm may imply the enhancement of another third order nonlinear tensor $\eta_{(3)}^{baba}$ (or $\eta_{(3)}^{abba}$) as a result of Berry curvature singularity in the vicinity of the Weyl nodes[35].

The CPGE response ($I_{1/\pi}$) is allowed in TaIrTe$_4$ by the crystal symmetry through the third order nonlinear tensor $\eta_{(3)}^{baba}$ ($\eta_{(3)}^{abba}$) under an in-plane $E_{DC}$ along crystallographic *a*(*b*)-axis (see Supplementary Information IV). Similar to the shift current response, besides the transition in the vicinity of the Weyl cones, many optical transitions can account for the $I_{1/\pi}$. However, we note the observed $I_{1/\pi}$ is probably dominated by the chiral response of the Type-II Weyl cones due to the following three reasons: First, the CPGE response at the vicinity of the Weyl nodes can be enhanced as a result of Berry curvature singularity at the vicinity of Weyl nodes, similar to the shift current response. Second, the $E_{DC}$ provided by the PTE effect is larger at the Weyl cones as the Seebeck coefficient is larger at the linear energy dispersion region, which has been experimentally verified on graphene, a linearly energy dispersed Dirac semimetal[36]. Third, the momentum scattering time $\tau$ in Eqn. (2) is longer for low excitation energy at the vicinity of Weyl cone compared to high energy levels as higher energy optical phonon scattering process may get involved and increase the momentum scattering rate of high energy carriers.

The CPGE response from the Weyl cone is a result of the chirality selection rule, which is illustrated in Fig. 4f. The optical transition from the lower part to the upper part of the Weyl cone is mutually determined by the chirality of the Weyl node and the excitation photon. For RCP excitation and $\chi$ =+1 WF, the optical transition is allowed on the +*k* side but forbidden on the −*k* side due to the conservation of angular momentum[30,37,38]. Due to the above selection rule, RCP will result in a unidirectional current along +*k* direction in $\chi$ =+1 WF, while LCP will result in a current with direction switched to -*k* direction. With a pair of WNs with opposite chirality and exactly the same band structure related by the crystal mirror symmetry (exactly the case in TaIrTe$_4$), the net response due to the above chiral selection rule should vanish. One possible route to observe the CPGE in a WSM relies on a tilted Weyl cone and the asymmetric photoexcitation at the nodal points due to the Pauli blocking effect [37], which has been identified experimentally in Type-I WSM TaAs [30]. In this work, the built-in electric field $E_{DC}$ plays a role in tilting the Fermi levels between opposite-chirality WNs after photoexcitation. Consequently, the imbalance of Fermi level will result in differently allowed momentum spaces for optical transitions at certain photon energy crossing the Weyl cone (see Supplementary Fig. S9b). For certain circular polarization (CP) excitation, due to the imbalance of the allowed momentum space for optical transitions, the generated current from the opposite WNs could not cancel each other and hence contribute to the net photocurrent that is transverse to the direction of $E_{DC}$ as defined by $\eta_{(3)}^{baab}$ and $\eta_{(3)}^{baba}$. Although the transverse current is perpendicular to $E_{DC}$, it is still detectable under our measurement scheme with its nonzero projection along the electrodes. Furthermore, if $E_{DC}$ switches sign, the generated $I_{1/\pi}$ should also switch the sign, which is the exact characteristic observed experimentally. The current direction of $I_{1/\pi}$ component switches sign as the measurement is shifting from red to blue spot when $E_{DC}$ switches sign (Fig. 4c). According to this result, the in-plane $E_{DC}$ provide additional control knob for chiral carriers beyond the light helicity.

At last, we note the Berry curvature enhanced giant shift current response has practical significance for highly sensitive photodetection applications, especially for technically important mid-infrared and terahertz region. The un-optimized maximum room temperature responsivity of 130.2 mA/W at 4 μm is ready to catch up the state of art MCT detector (600 mA/W) running at low temperature.[39,40] Such effect could potentially be much larger if second order instead of third order nonlinear effect is used with materials of suitable crystal symmetry. With doping control, the highly sensitive detection could apply over broad wavelength range especially for longer wavelength that is approaching the WN. On the other hand, the advantage of the third order nonlinear effect is the in-plane electric field which provides another convenient control over the chiral carrier injection in the Weyl cone beyond the light helicity, which opens up new experimental possibilities for studying and controlling the WFs and their associated quantum anomalies.

## Methods:
**Photo Response Measurement**
The laser sources used in the optical measurements were two quantum cascade lasers (QCLs) operating at 10.6 μm and 4 μm, and a He-Ne laser outputs continuum wave (CW) 633 nm. A 50X reflective objective lens was used to focus the beam on the sample. The optical spot sizes were about 40 μm, 10 μm and 1.5 μm for 10.6 μm, 4 μm and 633 nm, respectively, estimated from the scanning reflection image across a sharp edge. The sample was placed on a two-dimensional (x-y) stage to perform the spatial scanning. The current and reflection images were recorded at the same time, therefore, the absolute location of the photoinduced signal could be found by comparing to the simultaneously taken reflection image. The light was first polarized by a linear polarizer and then the polarization was further modulated using a rotatable quarter/half-wave plate characterized by an angle *θ* respective to the polarization direction of the linear polarizer. The schematic diagrams of the linear polarization dependent measurement by rotating a half-wave plate and the circular photo galvanic effect measurement by rotating a quarter-wave plate are shown in Supplementary Information XIII. Throughout the paper, we assign the light propagation direction as $-\hat{z}$, which is along the crystallographic c-axis that is perpendicular to the in-plane direction of the TaIrTe$_4$ flake.

## Crystal growth and device fabrication
TaIrTe$_4$ single crystals were synthesized by solid state reaction with the help of Te flux. All the used elements were stored and acquired in argon-filled glovebox with moisture and oxygen levels less than 0.1 ppm, and all manipulations were carried out in the glovebox. The elements of Ta powder (99.99 %), Ir powder (99.999 %), and Te lump (99.999 %) with an atomic ratio of Ta/Ir/Te =1 : 1 : 12, purchased from Sigma-Aldrich (Singapore), were loaded in a quartz tube and then flame-sealed under high-vacuum of $10^{-6}$ Torr. The quartz tube was placed in a tube furnace, slowly heated up to 1000 ℃ and maintained at this temperature for 100 h, and then allowed to cool to 600 ℃ at a rate of 0.8 ℃/h, followed by cooling down to room temperature. The shiny, needle-shaped TaIrTe4 single crystals could be obtained from the product. Freshly exfoliated thin TaIrTe4 flakes were transferred onto a 300 nm SiO$_2$/Si substrate. Standard electron-beam lithography technique was used to pattern electrodes, consisting of 5 nm Cr and 50 nm Au, on the TaIrTe4 samples to form devices.

**Numerical simulation of photoresponse**

To numerically calculate the photocurrent using DFT, the effect of $\boldsymbol{E}_{DC}$ on the second order optical response is added through a shifting of the Fermi level in the momentum space using the following relationship:

$$f(\boldsymbol{k}) = f^0\left(\boldsymbol{k} + \frac{e\tau \boldsymbol{E}_{DC}}{\hbar}\right) \qquad (2)$$

where $\tau$ is the momentum relaxation time, $f^0$ is thermal equilibrium distribution, and $\boldsymbol{k}$ is the crystal quasi-momentum. The existence of $\boldsymbol{a}$-axis orientating $\boldsymbol{E}_{DC}$ would break the mirror symmetry along $\boldsymbol{a}$-axis direction. In this way, we could calculate the effective third-order DC optical response (rank-4) tensors using:

$$\sigma_{(3)R}^{\alpha a\beta\gamma}(0,\omega,-\omega) = -\frac{i\pi e^3}{2\hbar^2}\int \frac{d\boldsymbol{k}}{8\pi^3}\sum_{n,m} f_{n,m}\left(r_{mn}^\beta r_{nm;\alpha}^\gamma + r_{mn}^\gamma r_{nm;\alpha}^\beta\right)\delta(\omega_{mn}-\omega) \qquad (3)$$

$$\bar{\sigma}_{(3)I}^{\alpha a\beta\gamma}(0,\omega,-\omega) = \frac{ie^3}{4\hbar^2}\int \frac{d\boldsymbol{k}}{8\pi^3}\sum_{n,m} f_{n,m}\left\{r_{mn}^\beta r_{nm;\alpha}^\gamma H_-(\omega_{mn},\omega) + r_{mn}^\gamma r_{nm;\alpha}^\beta H_-(\omega_{mn},-\omega)\right\} \qquad (4)$$

$$\tilde{\sigma}_{(3)I}^{\alpha a\beta\gamma}(0,\omega,-\omega) = \frac{ie^3}{8\hbar^2}\int \frac{d\boldsymbol{k}}{8\pi^3}\sum_{n,m} f_{n,m}\left(r_{nm}^\gamma r_{mn}^\beta - r_{nm}^\beta r_{mn}^\gamma\right)\frac{\partial H_-(\omega_{mn},\omega)}{\partial k^\alpha} \qquad (5)$$

Here, $\sigma_{(3)}^{\alpha a\beta\gamma} = \sigma_{(3)R}^{\alpha a\beta\gamma} + i\sigma_{(3)I}^{\alpha a\beta\gamma}$ and $\sigma_{(3)I}^{\alpha a\beta\gamma} = \bar{\sigma}_{(3)I}^{\alpha a\beta\gamma} + \tilde{\sigma}_{(3)I}^{\alpha a\beta\gamma}$. The index $\alpha, \beta, \gamma$ are dummy indices and can be any direction of crystal axes $\boldsymbol{a}$, $\boldsymbol{b}$ and $\boldsymbol{c}$. The second index $a$ is the direction of $\boldsymbol{E}_{DC}$, which is the $\boldsymbol{a}$-axis direction. The right-hand side of Eq.(3-5) is nothing but the definition of second-order shift current tensors with the fermi-factor $f_{n,m}$ modified by Eq.(2) to take into account of the effect of $\boldsymbol{a}$-axis orientating $\boldsymbol{E}_{DC}$ field. If $\boldsymbol{E}_{DC}=0$, the imaginary part $\sigma_{(3)I}^{\alpha a\beta\gamma}$ vanishes. $n$, $m$ are the band indices and $f_{n,m} = f_n - f_m$ is the distribution difference between $n$-th and $m$-th bands. All the effects of space charge DC field have been taken into account by the factor $f_{n,m}$. The factor $H_-(\omega_{mn},\omega) = \frac{P}{\omega_{mn}-\omega} - \frac{P}{\omega_{mn}+\omega}$ and $P$ is principal part. $r_{mn}^\alpha(\boldsymbol{k}) = \langle u_m(\boldsymbol{k})|i\partial_{k^\alpha}|u_n(\boldsymbol{k})\rangle$ measures the expectation value of the position operator in momentum space, and is nothing but Berry connection. $r_{mn;\beta}^\alpha$ is the generalized derivatives of $r_{mn}^\alpha$: $r_{mn;\beta}^\alpha = \frac{\partial r_{mn}^\alpha}{\partial k^\beta} - i\left[A_m^\beta(\boldsymbol{k}) - A_n^\beta(\boldsymbol{k})\right]r_{mn}^\alpha(\boldsymbol{k})$, and $A_m^\beta(\boldsymbol{k}) = \langle u_m(\boldsymbol{k})|i\partial_{k^\beta}|u_m(\boldsymbol{k})\rangle$ is the intraband Berry connection.

The shift currents induced by an incident optical field $\boldsymbol{E}_{opt}(t) = \boldsymbol{E}_{opt}(\omega)e^{-i\omega t} + \boldsymbol{E}_{opt}(-\omega)e^{i\omega t}$ with $\boldsymbol{E}_{opt}(-\omega) = \boldsymbol{E}_{opt}(\omega)^*$

$$J^{\alpha}_{\text{shift}} = \sigma^{\alpha a \beta \gamma}_{(3)}(0,\omega,-\omega)E^{\beta}(\omega)E^{\gamma}(-\omega) + \sigma^{\alpha a \beta \gamma}_{(3)}(0,-\omega,\omega)E^{\beta}(-\omega)E^{\gamma}(\omega) \qquad (6)$$

We consider the DC field is fixed along crystallographic **a**-axis. Non-zero planar components of $\sigma^{aaaa}_{(3)}$, $\sigma^{aabb}_{(3)}$, $\sigma^{baab}_{(3)}$, and $\sigma^{baba}_{(3)}$. According to Eq.(3-5) $\sigma^{\alpha a \beta \gamma}_{(3)R}$ is symmetric with respect to the interchange of the last two Cartesian components while $\sigma^{\alpha a \beta \gamma}_{(3)I}$ is antisymmetric. Here $\sigma^{aaaa}_{(3)}$ and $\sigma^{aabb}_{(3)}$ account for the longitudinal shift current, while $\sigma^{baab}_{(3)}$ and $\sigma^{baba}_{(3)}$ account for the transverse shift current.

The injection current tensor can be computed as:

$$\eta^{\alpha a \beta \gamma}_{(3)}(0,\omega,-\omega) = \frac{\pi e^3}{2\hbar^2}\int \frac{d\mathbf{k}}{8\pi^3}\sum_{n,m}\Delta^{\alpha}_{mn}f_{nm}\left(r^{\gamma}_{mn}r^{\beta}_{nm} - r^{\beta}_{mn}r^{\gamma}_{nm}\right)\delta(\omega_{mn}-\omega). \qquad (7)$$

The right-hand side of Eq.(7) is nothing but the definition of second-order injection current tensor with the fermi-factor $f_{n,m}$ modified by Eq.(2) to take into account of the effect of **a**-axis orientating $E_{\text{DC}}$ field. Here $\Delta^{\alpha}_{mn} = v^{\alpha}_{m} - v^{\alpha}_{n}$, and $v^{\alpha}_{m} = \left\langle u_m(\mathbf{k})\left|\frac{\partial H_k}{\hbar \partial k^{\alpha}}\right|u_m(\mathbf{k})\right\rangle$ is the velocity. $\eta^{\alpha a \beta \gamma}_{(3)}$ is purely imaginary and antisymmetric with respect to the interchange of the last two Cartesian components, $\eta^{\alpha a \beta \gamma}_{(3)} = -\eta^{\alpha a \gamma \beta}_{(3)}$. The injection current only has the transverse component.

## Data availability

The data that support the plots within this paper and other findings of this study are available from the corresponding author upon reasonable request.

## Acknowledgement

The authors acknowledge helpful discussions with Qiong Ma. This project has been supported by the National Natural Science Foundation of China (NSFC Grant Nos. 91750109, 11674013, 11774010, 11704012, 11374021), the National Basic Research Program of China (973 Grant No. 2014CB920900), the National Key Research and Development Program of China (Grant Nos:2018YFA0305604), the Recruitment Program of Global Experts and the State Key Laboratory of Precision Measurement Technology and Instruments Fund for open topics.

## Author Contribution

D.S. conceived the idea and designed the experiments. Y.P. synthesize the bulk TaIrTe$_4$ materials under the supervision of Z.L.. Y.N.L. fabricated the TaIrTe$_4$ devices under the supervision of J.H.C.; J.C.M., J.W.L. and X.Z. performed the optical measurements under the supervision of D.S.; Q.Q.G. performed the band calculations under the supervision of J.F.; J.C.M., Q.Q.Gu, J.F. and D.S. analyzed the results; D.S. wrote the manuscript, assisted by J.C.M., Q.Q.Gu, J.H.C., and J.F. All the authors comment on the manuscript.

## Additional Information

**Supplementary Information** accompanies this paper at http:/www.nature.com/naturecommunications

**Competing Interest:** The authors declare no competing financial interest